\documentclass{article} 
\usepackage{graphicx} 
\usepackage{amsmath}
\usepackage{siunitx}
\usepackage{float}
\usepackage{subcaption}
\usepackage[a4paper, margin=2.5cm]{geometry}
\usepackage{subcaption}
\usepackage{booktabs}
\usepackage{multirow}
\usepackage[dvipsnames]{xcolor}
\usepackage{bm}
\usepackage{cite}
\usepackage{comment}
\usepackage{hyperref}

\title{3D Plasma plume characterization of an electrodeless thruster cluster in magnetic arch configuration}
\author{Sacha A. Huot, Marco Inchingolo, Jaume Navarro-Cavallé, Mario Merino}
\date{November 2025}

\begin{document}

\maketitle

\begin{abstract}
Clustering electrodeless plasma thrusters in pairs with opposing magnetic polarities offers an easy means to scale-up the propulsion system of future missions, and also, to mutually cancel their respective magnetic dipoles. Their magnetic nozzles merges to form a new topology, the `magnetic arch', which can yield a lower plasma plume divergence than two separate magnetic nozzles. 
This work characterizes the plasma expansion in the fully-closed magnetic arch of a cluster of two electron-cyclotron resonance thrusters with electrostatic probes (Langmuir probes, Faraday cups, and a Retarding Potential Analyzer).
Electrostatic potential, plasma density, electron temperature, ion current and energy are measured in the two orthogonal planes of symmetry of the setup for various operating conditions.
Results show that a plasma jet can be extracted even from this magnetic configuration, albeit with a reduced ion energy. 
A slight potential hill and hotter electrons exist in the central part of the arch.
Ion current profiles are doubly-peaked in the horizontal plane, likely corresponding to the beamlet of each thruster. 
Trends with xenon mass flow rate and input power are consistent with the expectations of electrodeless plasma thrusters.
The plume experiences an upward or downward deflection depending on the direction of the applied magnetic field, which could be attributed to the effect of the lateral electron drifts in the magnetic arch.
\end{abstract}

\section{Introduction}

Electric propulsion is widely used in spacecraft applications due to its high specific impulse compared to chemical propulsion systems. Among the different propulsive concepts, electrodeless plasma thrusters (EPTs) have attracted increasing attention as a promising alternative, as they do not rely on electrodes, suppressing the erosion challenge, can operate with virtually any propellant, and can potentially offer long operational lifetimes.
Two prominent EPT concepts are helicon plasma thrusters (HPTs) \cite{BoswellHPT2,navarro-cavalleExperimentalCharacterization12018,TakahashiHPT} and electron cyclotron resonance thrusters (ECRTs) \cite{MillerECRT,Sercel_Culick_1993,cannat2015optimization,inchingoloPlumeCharacterizationWaveguide2023,ECRperformanceSheppard}. 

In these devices, plasma is generated and heated by radio-frequency or microwave electromagnetic waves and subsequently accelerated through a magnetic nozzle, which converts the electron thermal energy into directed ion momentum \cite{ahedo2010, ahedo2011magnetic, TakahashiMN}. The magnetic nozzle therefore plays a central role in shaping the plume divergence and controlling the energy conversion efficiency of the plasma expansion. Over the past decades, several experimental  \cite{navarro-cavalleExperimentalCharacterization12018, littleIonizationCurrentSheet2022, cannat2015optimization, inchingoloPlumeCharacterizationWaveguide2023}
and numerical 
\cite{svil21a, svil23,jime23}
investigations have explored the performance and plume characteristics of EPTs, with ECRT technology in particular having been developed and studied by several research groups. These studies have examined plasma detachment, plume divergence, and energy conversion processes in magnetic nozzles, highlighting the complex coupling between plasma dynamics and magnetic topology.

In the case of a single EPT, two practical challenges arise from its design. First, unless properly compensated for, a net magnetic dipole (intrinsic to the magnetic nozzle), leads to interactions with planetary magnetic fields. This dipole can be either temporary in the case of purely electromagnet-generated nozzles, or constant if permanent magnets are employed, with a direct impact on the attitude control system when there is a background magnetic field, especially in low Earth orbit. The second issue is the relatively large plume divergence observed in both numerical simulations and experimental works \cite{ahedo2010,cannat2015optimization,inchingoloPlumeCharacterizationWaveguide2023}. Together, these two aspects can hinder the spacecraft integration of single EPTs.

One potential solution to mitigate these drawbacks is the clustering of pairs of EPTs. Appropriate magnetic configurations within a cluster can modify the magnetic topology of the plume. In particular, pairing two thrusters with opposite magnetic polarity generates a \textit{magnetic arch} connecting the two plasma sources. This arch can be fully-closed, when all magnetic field lines connect the two sources, or partially-open if some outer field lines do not connect directly and form a separatrix between open and closed regions.
Two advantages arise from this configuration. The first is the cancellation of the net magnetic dipole, yielding a cluster with zero net dipole moment. The second is the potential reduction of plume divergence due to the confining topology created by the magnetic arch.
In addition, clustering allows the propulsion system to scale up in power and introduce some degree of thrust vectorability in a straightforward manner while introducing redundancy, which can improve reliability.

Plasma expansion in a magnetic arch presents significantly different physics compared with expansion in a conventional  magnetic nozzle. In a magnetic arch, part of the magnetic topology consists of closed field lines connecting the two sources, while other field lines may remain open and guide the plasma downstream. This topology causes the interaction of the two plasma streams and modifies electron confinement, ion acceleration mechanisms, and plume structure.


The clustering of two HPTs was experimentally investigated by Vereen et al. \cite{vereenCharacterizationClusterHigh2017}. The cluster produced a plume structure that was noticeably different and achieved larger ion currents densities than the simple superposition of two independent plumes. However, in that work, only two HPTs with the same magnetic polarity were operated simultaneously, a concept completely different that of the magnetic arch.
The first experimental characterization of a cluster operating in a magnetic arch configuration was carried out by Boyé et al. \cite{Boyé_Navarro-Cavallé_Merino_2025}, where two ECRTs with opposite polarity were combined. That work measured the ion current and energy in a limited region of the fundamental plane of the magnetic arch (what we refer to as the `horizontal' plane) and demonstrated that plasma could be extracted from a partially-open configuration, producing a plume with reduced divergence compared with single-source operation, although with a 30 \% reduction in ion energy with respect to a single magnetic nozzle. 
Nevertheless, experimental measurements of the plasma properties outside this primary symmetry plane remain pending.

These findings have also been supported by numerical investigations. Simulations using planar fluid models \cite{merinoPlasmaAccelerationMagnetic2023} and hybrid PIC/fluid approaches \cite{Guaita_2025} confirmed that plasma beams can indeed be extracted from magnetic arch configurations. These studies revealed that the complex interaction between the two beamlets generated by each source. In particular, the simulations predict the formation of a potential hill in the interaction region where plasma density increases. This potential structure can trap part of the electron population along closed magnetic field lines while guiding ions along open field lines toward the plume region. It is also responsible for the lower ion energy registered downstream.
However, both these numerical and experimental investigations have focused on two-dimensional configurations, typically assuming symmetry with respect to the plane containing the two sources. As a result, they do not capture the plasma response outside this fundamental symmetry plane of the cluster.
Full three-dimensional scaled-down simulations of clustered helicon plasma thrusters were carried out by Di Fede et al. \cite{difedeMagneticNozzlePerformance2023}. While their analysis focuses mostly on the horizontal symmetry plane, their results do not suggest additional large-scale three-dimensional plasma dynamics.

In this work, we perform the first full characterization of a cluster of ECRTs in a fully-closed magnetic arch configuration by performing a detailed characterization of the plasma plume in the two orthogonal symmetry planes (i.e., `horizontal' and `vertical' planes)  using electrostatic probes. Faraday cup, retarding potential analyzer (RPA), and Langmuir probe measurements are employed to characterize ion flux, ion energy distributions, and local plasma parameters in the plume.
The experimental setup used in the previous work of Boyé et al. \cite{Boyé_Navarro-Cavallé_Merino_2025} has been modified to produce a strictly fully-closed magnetic arch configuration, with the intention of emphasizing the physical mechanisms that are specific to magnetic arch topologies as opposed to conventional open magnetic nozzle expansions. A companion work by Boyé et al \cite{boye:hal-04686118} characterizes the ion velocity distribution function in this setup using laser induced fluorescence spectroscopy.
The main findings of the present study show that the plasma plume exhibits a structured topology strongly shaped by the magnetic arch, shown by extensive measurements in the horizontal plane. Measurements in the vertical plane reveal additional information on the three-dimensional structure of the plume, confirming that the magnetic arch strongly influences plasma confinement and ion beam expansion beyond the primary horizontal plane. Plasma structure is encountered to be not symmetric within the vertical plane, which is, at first glance, controversial with respect to the apparently double symmetry of the cluster setup. 

The reminder of this article is structured as follows.
Section \ref{sec:experimentalsetup} details the experimental setup, the test facility, and the plasma diagnostics used.
Section \ref{sec:results} presents and discusses the electrostatic probes measurement results.
Finally, Section \ref{sec:conclu} gathers the main conclusions of this work.

\section{Experimental setup}\label{sec:experimentalsetup}

\subsection{ECRT Cluster}

\begin{figure}
    \centering
    \includegraphics[width=\linewidth]{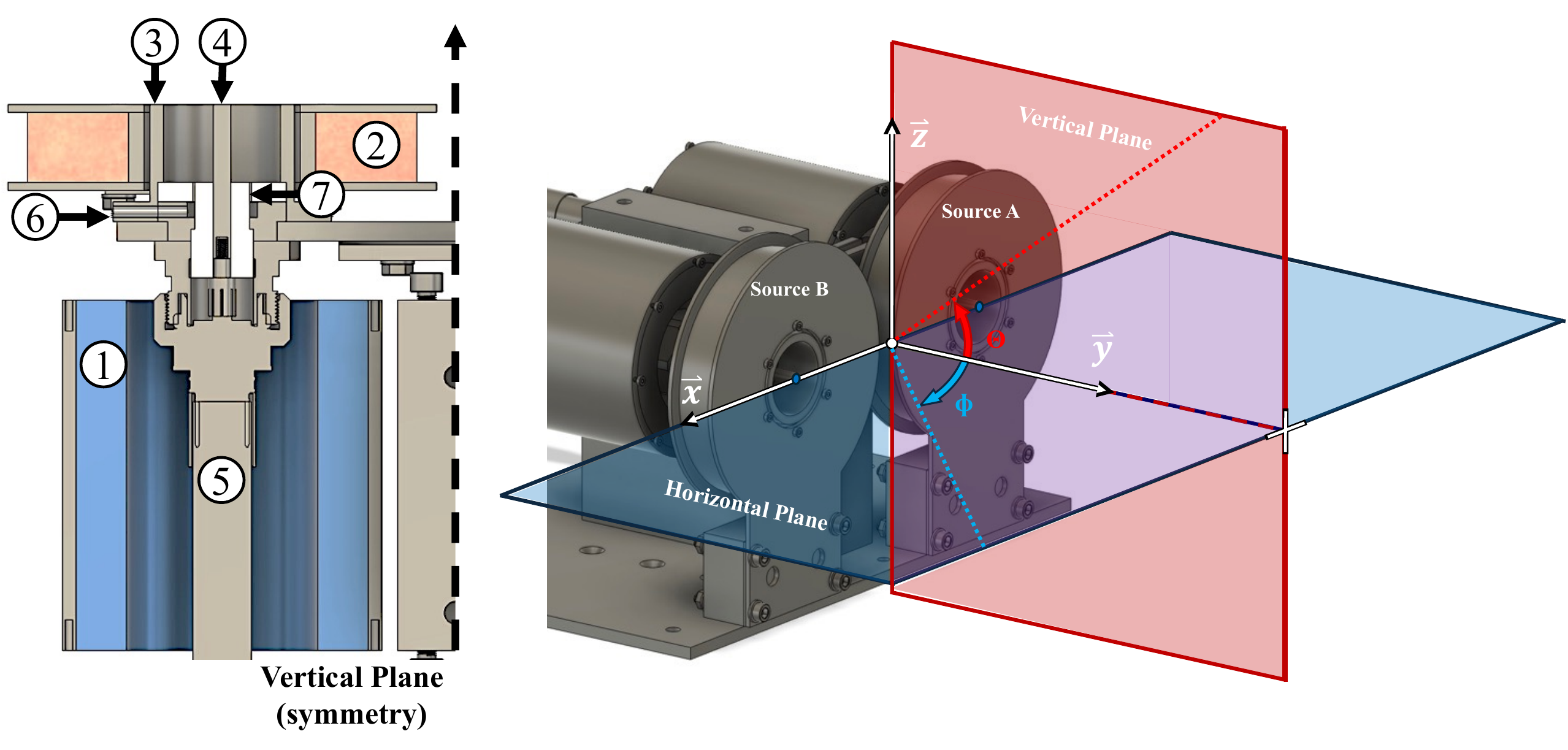}
    \caption{ECRT cluster's section view (left) showing the permanent magnets (1), electromagnetic coil (2), plasma chamber (3), central conductor (4), coaxial cable (5), gas inlet (6), circular ceramics (7). (right) CAD isometric view of the ECRT cluster setup, with the representation of the horizontal and vertical planes, in which measurements are taken. Angles $\Phi, \Theta$, unambiguously determine the angular position of the electrostatic probes.}
    \label{fig:CAD}
\end{figure}

The setup used in this work is based on the one used in \cite{Boyé_Navarro-Cavallé_Merino_2025}.
The Magnetic Arch Cluster consists in a cluster of two ECR sources mounted side by side and parallel to each other. Figure \ref{fig:CAD} shows a CAD view of the assembly. In each source, the magnetic field is generated by nine N42 permanent ring magnets (1) mounted together behind the plasma sources and a coil or electromagnet (2) placed around each plasma chamber (3). The plasma sources are cylindrical chambers (15 mm radius, 20 mm depth), along whose axis lays the central conductor (4) of the coaxial microwave transmission line (5) carrying the alternating electromagnetic field. The plasma chambers are connected to the coaxial microwave transmission line by standard DIN 7/16 connectors. Each plasma source forms the termination of a coaxial line: the cylindrical chamber acts as the outer conductor, while the central conductor extends along the chamber axis. The chamber and central conductor are electrically connected with the outer and inner conductors of the transmission line, respectively. The propellant gas (6) is axially injected into the chambers through an annular gap between the ceramic backplates (7) separating the chamber from its coaxial connector. 
The key differences with the setup used in \cite{Boyé_Navarro-Cavallé_Merino_2025} are the shorter length of the plasma chambers (20 mm instead of 43.7 mm) and the use of permanent magnets behind the chambers instead of an electromagnet, leading to the difference in magnetic arch topology.

Xenon gas is fed through a single mass flow controller (MFC), into a single line to the side of both plasma chambers through Teflon tube and a three way splitter. From the splitter, it reaches the annular gap described above by means of a radial line, made of stainless steel, on the side of the chamber.
Along the experimental campaign, the total xenon mass flow rate $\dot{m}$ was varied from 5 to 15 sccm.

\begin{figure}
    \centering
    \includegraphics[width=\linewidth]{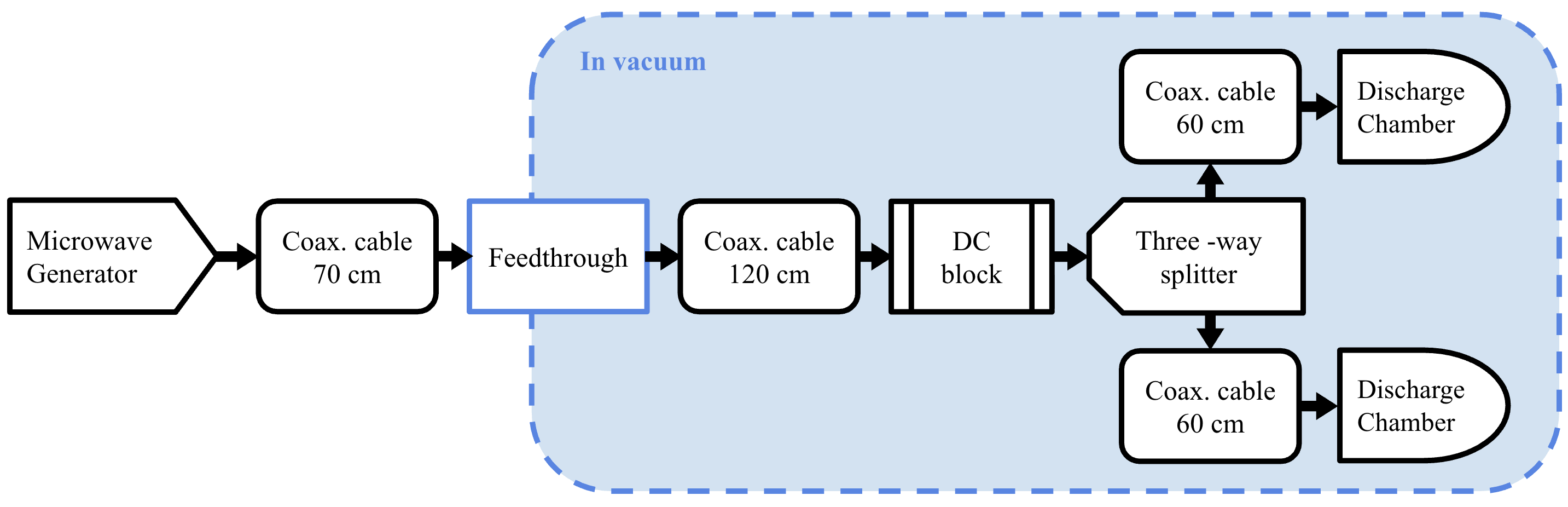}
    \caption{Transmission line diagram.}
    \label{fig:TL}
\end{figure}

The microwaves are generated by a single 2.45 GHz solid state microwave generator, model MR1000D-200ML from \textit{Muegge} and transmitted to the plasma chambers by a coaxial line, schematically represented on figure \ref{fig:TL}. Standard DIN 7/16 coaxial cables are used and a three way splitter allows to separate the line in two terminations to be connected to each source. As the cluster needs to remain electrically floating, a DC-block is placed in the transmission line between the generator and the three-way splitter in order to insulate the cluster while transmitting the microwaves. The DC block is a set of two coaxial to waveguide adapters (DIN 7/16 to WR430) facing each other and separated by 1 mm ceramic wafers to ensure insulation and alignment. Each component has been tested for transmission and reflection coefficients with a virtual network analyzer (VNA) to guarantee the lowest amount of microwave reflection possible in the transmission line.

From the VNA measurements, the power transmitted is: 36,43 \% to source A, 36,70 \% to source B, i.e. a total transmitted power of 73.33 \%. The main contributor to losses were identified to be the long (120cm \& 70cm) and bended coaxial cables, used in our experimental setup for convenience.
In order to avoid the potential overheating of components during the experimental campaign, the temperature of the following elements was monitored by thermocouples: feedthrough, three-way splitter, DC block, discharge chamber (source A \& B backside where the coaxial cable ends). 
In these experiments, forward power $P$ was varied between 50 and 150 W.

\begin{figure}[H]
    \centering

    \begin{subfigure}{.49\linewidth}
        \centering
        \includegraphics[width=\linewidth]{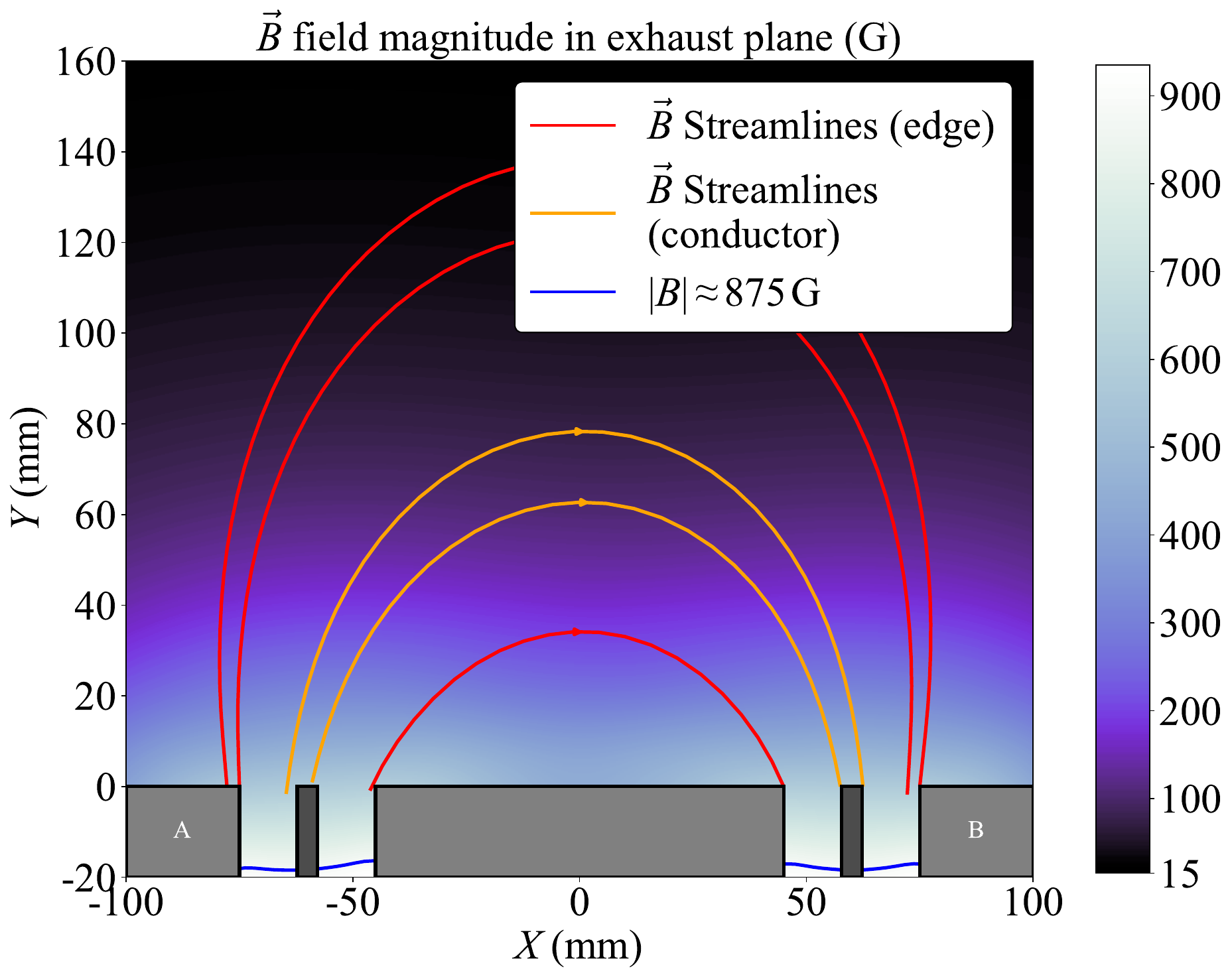}
        \subcaption{}
    \end{subfigure}
    \hfill
    \begin{subfigure}{.49\linewidth}
        \centering
        \includegraphics[width=\linewidth]{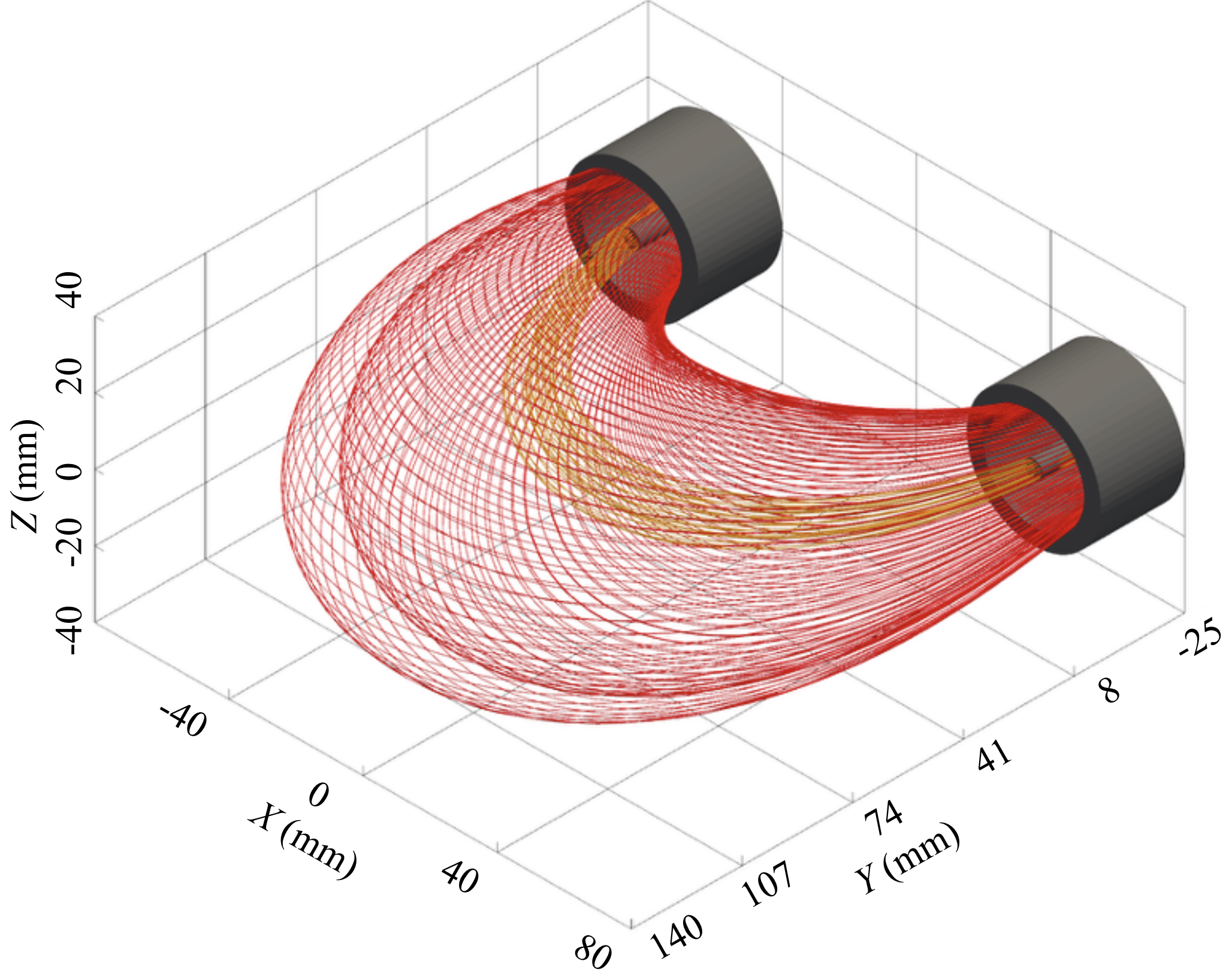}
        \subcaption{}
    \end{subfigure}

    \vspace{0.5cm}

    \begin{subfigure}{.85\linewidth}
        \centering
        \includegraphics[width=\linewidth]{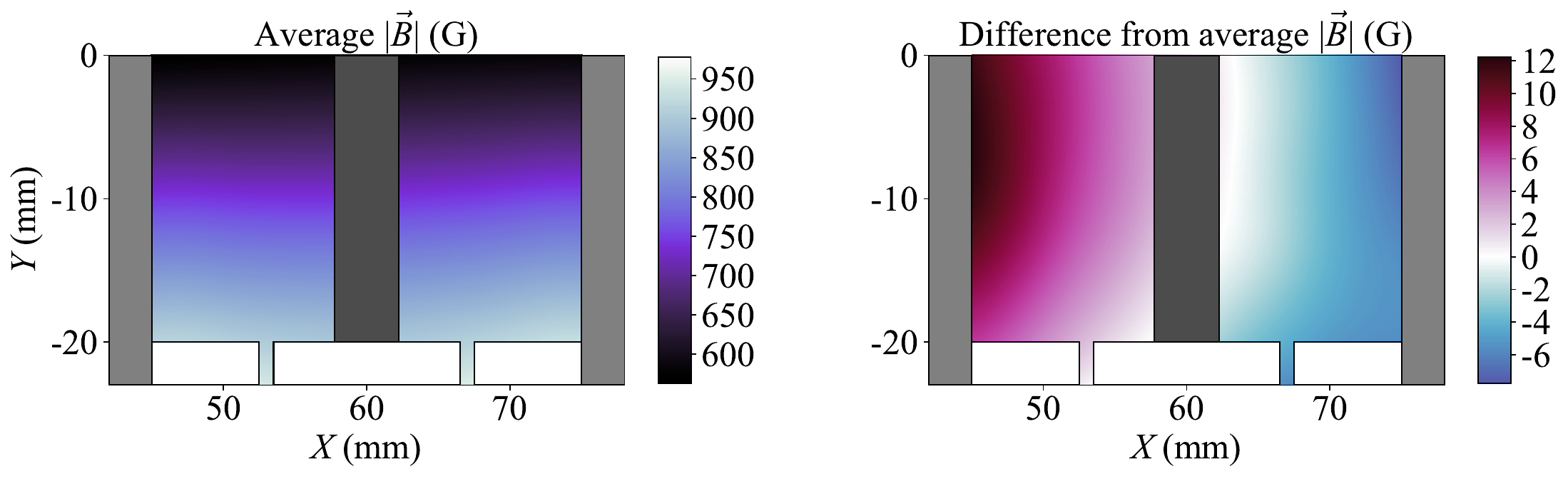}
        \subcaption{}
    \end{subfigure}

    \caption{Magnetic arch field intensity and main streamlines in the horizontal plane (a), 3D streamlines originating form the chambers and central conductor edges (b). Magnetic field asymmetry inside the discharge chambers (c), the difference from the average field is shown for source B. 
    }
    \label{fig:Bfield}
\end{figure}


The magnetic field of the arch was characterized experimentally prior to plasma operation using a three-axis gaussmeter, model 460 3-channels Gaussmeter from \textit{Lake Shore Cryotronics}, and is presented on figure \ref{fig:Bfield}(a-c). Measurements were performed in air on a three-dimensional grid covering the inside of the discharge chambers and the exhaust plume volume, using an automated arm system with three degrees of freedom. Two measurements were performed, one with the coils turned off, and one with 2 A of current in each coil with their polarity aligned with their respective permanent magnet. The latter is the nominal magnetic field configuration used in this work, and the only one presented here. A thinner measurement mesh was used in regions of stronger magnetic field gradients i.e. the discharge chambers where the ECR heating occurs and close plume region. The measured magnetic field strength was used to identify the ECR resonance surfaces corresponding to the microwave frequency employed in this work.

At 2.45~GHz, the electron cyclotron resonance occurs at approximately 875~G. By design,the ECR surfaces are located deep inside each discharge chamber, near the ceramic backplate of the sources.
Downstream of the chambers, the magnetic field strength decreases rapidly and the streamlines originating from the resonance regions extend into the near plume to form the magnetic arch connecting the two sources.

Magnetic field lines traced from the measured field maps are shown in figure \ref{fig:Bfield}(a,b). The streamlines originating from the central conductor edges define the core of the magnetic arch (orange), while streamlines originating from the chamber edges delineate its outer boundary (red). This topology results in a closed magnetic flux tube in the near plume region. The structuring effect of the flux tubes on the plasma plume will be detailed in section \ref{sec:results}.
Within this arch, the magnetic field strength varies from 570 G at the exit of the sources to 25 G at the most downstream point.

An asymmetry between the field strength inside of the sources is noted, as shown on figure \ref{fig:Bfield}(c). There is a small ($\simeq 10$G) difference in field intensity with respect to the average field inside the chambers on the inner most side of the sources. This small difference, attributed to defects in the permanent magnets, is the reason for the imperfect overlap of the outer streamlines in red on figure \ref{fig:Bfield}(a,b).


No optimization of the individual sources was done in terms of their propulsive figures of merit for this experiment, which focuses on understanding the main physics in the interaction of the magnetized plumes; indeed, the sources provide only mild-energy ion jets and exhibit limited propellant utilization \cite{Boyé_Navarro-Cavallé_Merino_2025}.

\subsection{Test facility and diagnostics}

The magnetic arch cluster was tested in the large vacuum chamber of the EP2 laboratory at the Universidad Carlos III de Madrid (3.5 m long, 1.5m in internal diameter).
Pumping is achieved through a staged configuration consisting of a Leyvac~LV80 roughing pump ($22.3\,\mathrm{L\,s^{-1}}$), two MAGW2.200iP turbomolecular pumps (each rated at $2000\,\mathrm{L\,s^{-1}}$), and a set of Leyvac~140~T-V in combination with 3 cryopanels.
The vacuum facility provides a total effective pumping capacity of approximately $3.7\times10^{4}\,\mathrm{L\,s^{-1}}$, enabling an ultimate base pressure below $10^{-7}\,\mathrm{mbar}$ and stable continuous operation pressures below $2\times10^{-5}\,\mathrm{mbar}$. 
During cluster operations, the pressure is stable in the range of $<2\cdot10^{-5}$ mbar, for a Xenon mass flow rate of 5 to 15 sccm.

The plasma diagnostics apparatus used to characterize the magnetic field and various plasma parameters include Langmuir probes, a Faraday cup, and a RPA. 
The probes are mounted on a 2 degrees of freedom robotic arm system, placed downstream of the cluster inside the vacuum chamber. The arm is capable of azimuthal and radial motion in the horizontal plane, at a fixed height setting. The azimuthal motion has a range of $\theta = \pm 90 ^\circ $ and the radial motion has a range of $r \in [0,400]$ mm, in increments of $1^\circ $ and 1 mm respectively. The angular range is restricted in the actual measurements to avoid collision with the setup. Probes are aligned with respect to the origin of the measurement reference frame, located at the midpoint between the two sources.  

The Langmuir probe, built in-house \cite{perr24b}, features an electrode 3 mm in length and 0.254 mm in diameter.
The probe bias voltage is controlled by a high resistance electrometer, model 6517B from \textit{Keithley} with a current resolution of 1 nA and voltage resolution of 50 mV. The acquisition of the $I-V$ curve is automated via a LabVIEW program.
The $I-V$ curves obtained using Langmuir probes are preprocessed using the method described by Lobbia and Beal in \cite{Lobbia_Beal_2017} to extract the plasma parameters, over the full range of radii that can be scanned by the arm system.
The Faraday cup, also built in-house \cite{inch23b}, is equipped with a circular aperture with a radius of 5 mm. The ion current is measured by a high resistance electrometer, model 6517B from \textit{Keithley}.
Faraday cup measurements were taken on an arc at $r=280$ mm distance from the origin of the reference frame.
The RPA is the model Semion RFEA from \textit{Impedans}, same as the one used in previous works \cite{inch24a,boye25a}. A radius of $r=380$ mm was used for the RPA measurements.
 
\section{Results}\label{sec:results}
%

\begin{figure}[H]
    \centering

    \begin{subfigure}{0.49\linewidth}
        \centering
        \includegraphics[width=\linewidth]{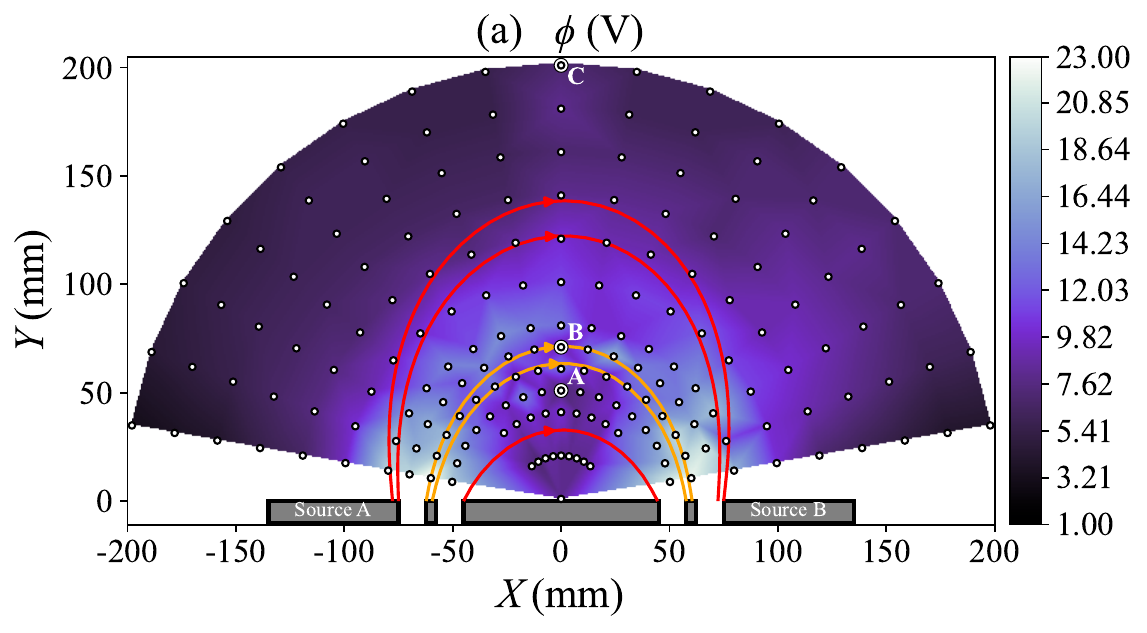}
        \label{fig:VP_ref_H}
    \end{subfigure}
    \hfill
    \begin{subfigure}{0.49\linewidth}
        \centering
        \includegraphics[width=\linewidth]{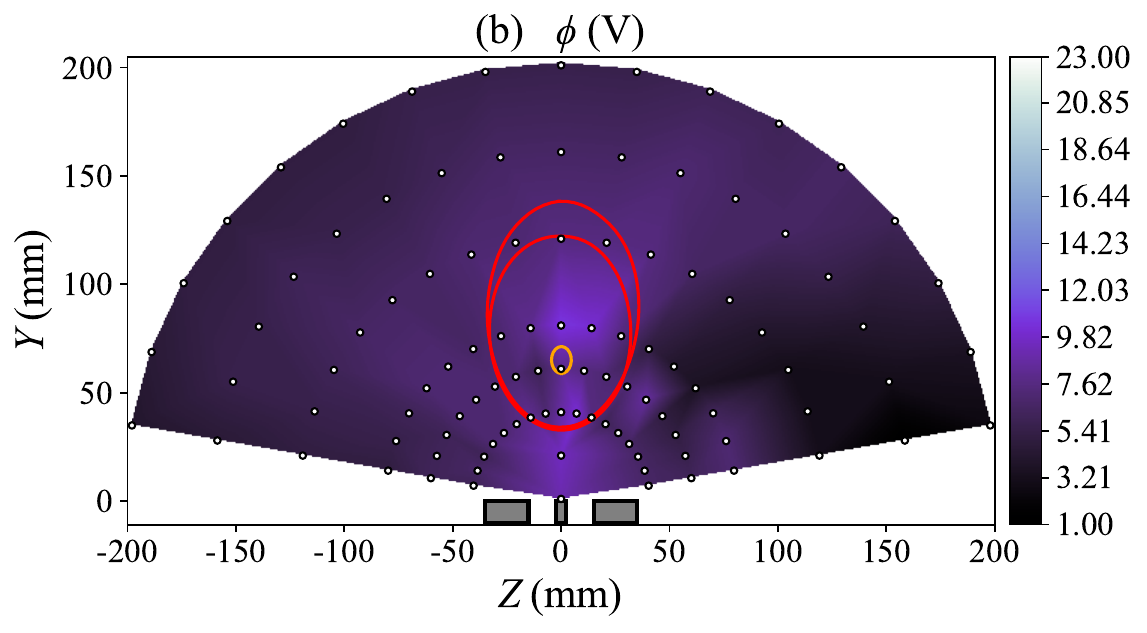}
        \label{fig:VP_ref_V}
    \end{subfigure}
    \begin{subfigure}{0.49\linewidth}
        \centering
        \includegraphics[width=\linewidth]{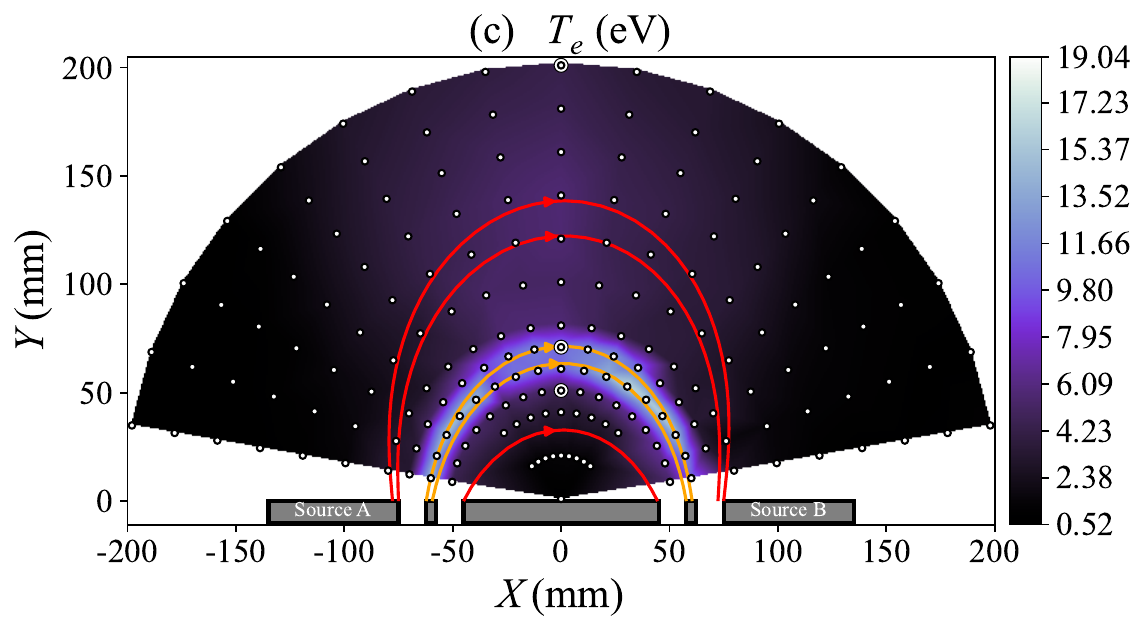}
        \label{fig:Te_ref_H}
    \end{subfigure}
    \hfill
    \begin{subfigure}{0.49\linewidth}
        \centering
        \includegraphics[width=\linewidth]{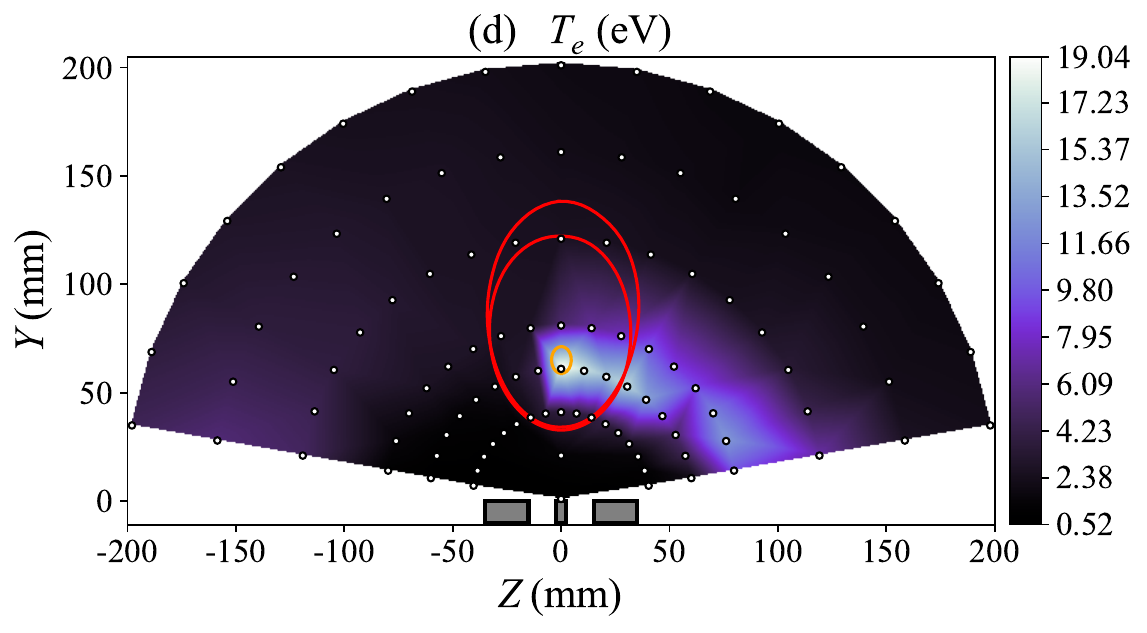}
        \label{fig:Te_ref_V}
    \end{subfigure}
    \begin{subfigure}{0.49\linewidth}
        \centering
        \includegraphics[width=\linewidth]{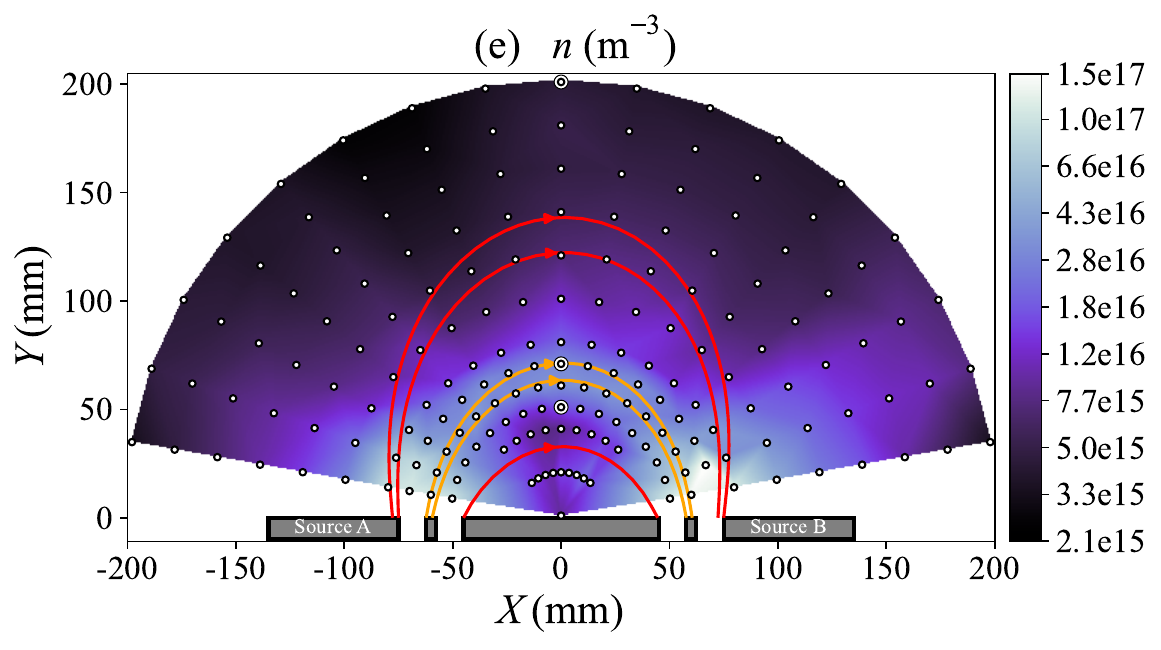}
        \label{fig:Ni_ref_H}
    \end{subfigure}
    \hfill
    \begin{subfigure}{0.49\linewidth}
        \centering
        \includegraphics[width=\linewidth]{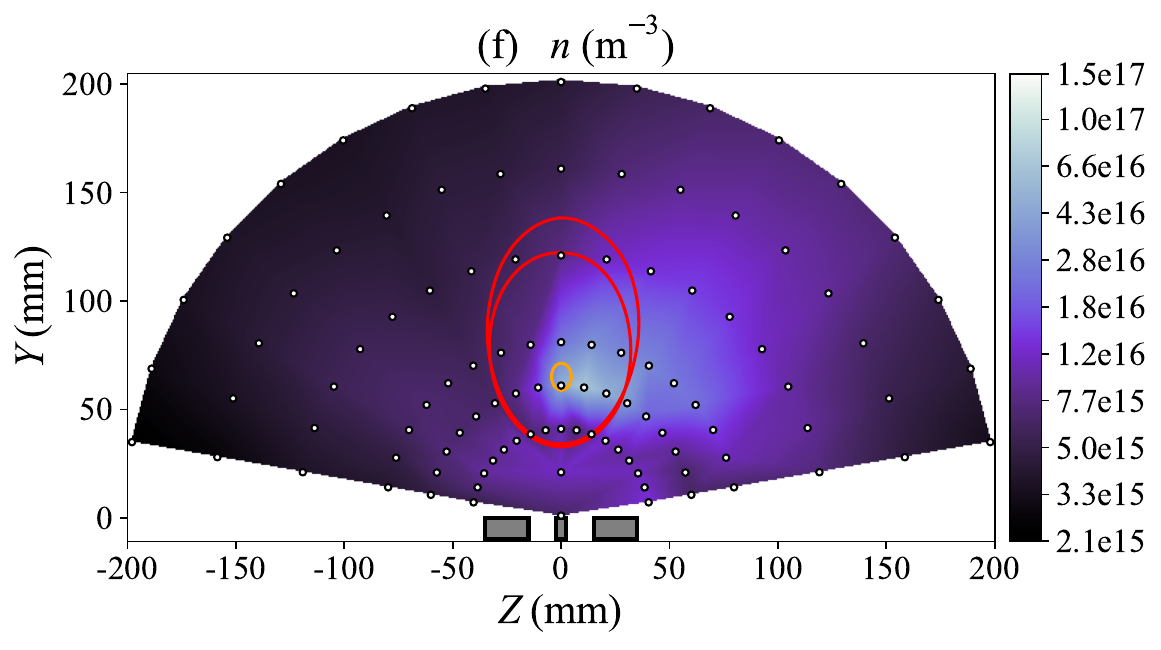}
        \label{fig:Ni_ref_V}
    \end{subfigure}

    \caption{
    Langmuir probe measurements for the reference case ($P=100$ W,
    $\dot{m}=10$ sccm).
    Left column: horizontal plane.
    Right column: vertical plane.
    Plasma potential (a,b), electron temperature (c,d), and plasma density (e,f).
    Measurement points are indicated by white dots. Main magnetic streamlines or
    arch intersection are shown in red (lines passing through the chamber edge) and orange (passing through the inner conductor edge). Points of interests are highlighted and named $A$, $B$, $C$.}
    \label{fig:LP_reference_combined}
\end{figure}

Langmuir probe measurements were used to characterize the plasma potential $\phi$, electron temperature $T_e$, and plasma density $n$ in the plume of the ECRT cluster. Scans were performed in both the horizontal and vertical planes (see figure \ref{fig:CAD}) for three operating points ($P=100$~W, $\dot{m}=5, 10, 15$~sccm). The resulting maps for $\dot{m}=10$~sccm are shown in figure ~\ref{fig:LP_reference_combined}.

In the horizontal plane, the electrostatic potential $\phi$ reaches its maximum in the immediate vicinity of each plasma chamber, with peak values of approximately 22~V. These mild values are in close agreement with the most probable ion energy of approximately 24~eV measured by the RPA as discussed further below. Furthermore, they are compatible with the ion velocity values reported by Boyé et al. in \cite{boye:hal-04686118} using laser induced fluorescence on the same experimental configuration: $P=100$~W, $\dot{m}=5$~sccm.

The electrostatic potential decreases initially away from each source, resembling the behavior in a single MN.
However, it soon plateaus into a diffuse region, that we term the `magnetic arch tube'. This is roughly delimited by the red magnetic streamlines born at the lips of the ionization chamber of each source; the slight asymmetry in the magnetic field of each source makes the outermost lines to not coincide.
This arch tube is marked by a sustained plasma potential (around 15~V), indicating that this electrostatic potential structure is influenced by the magnetic topology, and therefore governed by the magnetized electrons, attached to the lines. In the inner region between the two sources 
upstream of the arch tube, the plasma potential shows lower values (around 8~V), similar to what can be measured in the outer region beyond the arch tube. There, the potential decreases monotonically downstream, again resuming the behavior of the plasma expansion in a single MN (and any general plasma expansion to vacuum).

In this horizontal plane, the electron temperature map exhibits a marked spatial structure. A population of warmer electrons is confined within the `core tube' of the magnetic arch tube. This is the magnetic tube delimited by the orange magnetic streamlines that meet the inner conductor element of each source, which protrude roughly 40--60~mm from the exit plane. Peak temperatures in this tube reach approximately 15~eV. 
Interestingly, this region of high electron temperature is much narrower than the region of high $\phi$, which roughly matches the whole arch tube.

Narrow tubes of high electron temperature are known to exist in single-ECRTs, which show stronger electron heating by the wave near the antenna, where the MW field is the strongest. This was also observed by ONERA on their ECRT \cite{peterschmittImpactMicrowaveCoupling2021}, and was also predicted by simulation works carried by Sanchez-Villar et al. \cite{svil21a}.
In our case, the high-$T_e$ tube is curved, clearly following the geometry of the magnetic field lines, which attests to the electrons being well-magnetized in the arch, over several centimeters downstream of the cluster. The electron magnetization is confirmed by their Larmor radius: from 0.6 mm at the exit plane to 5 mm at the edge of the arch, 140 mm downstream of the cluster.

The presence of electrons with energies exceeding the first ionization threshold of xenon (12.13~eV) suggests that these electrons are capable of ionizing neutral xenon escaping the plasma chambers, which feature incomplete propellant utilization. Thus, ionization likely occurs not only inside the sources, but in this core magnetic arch region. In addition, charge exchange collisions between neutrals and ions can take place in this region. This is consistent with the relatively broad ion energy distributions displayed by the RPA on figure \ref{fig:RPA all}, as ions generated downstream inside the arch experience a lower potential drop than those produced in the chambers, leading to energy spread.
An efficient ECRT source design with greater propellant utilization would likely limit this effect.
Downstream from the magnetic arch, the horizontal scan reveals a conical region of moderate $T_e$ (typically 5--8~eV) expanding into the plume.
 
The plasma density map features higher density  close to the exit of the plasma chambers as expected with individually-operating sources, but also in the core magnetic arch tube described above. 
This complex map 
is interpreted as the superposition of the contribution of primary ions emanating from the plasma sources, and a non-negligible plasma production in the hot-electron core of the arch, as suggested before. A majority of the low-energy electrons created in the ionization events in this region will remain confined by the potential, giving rise to a trapped electron population.

\begin{figure}
    \centering
    \includegraphics[width=\linewidth]{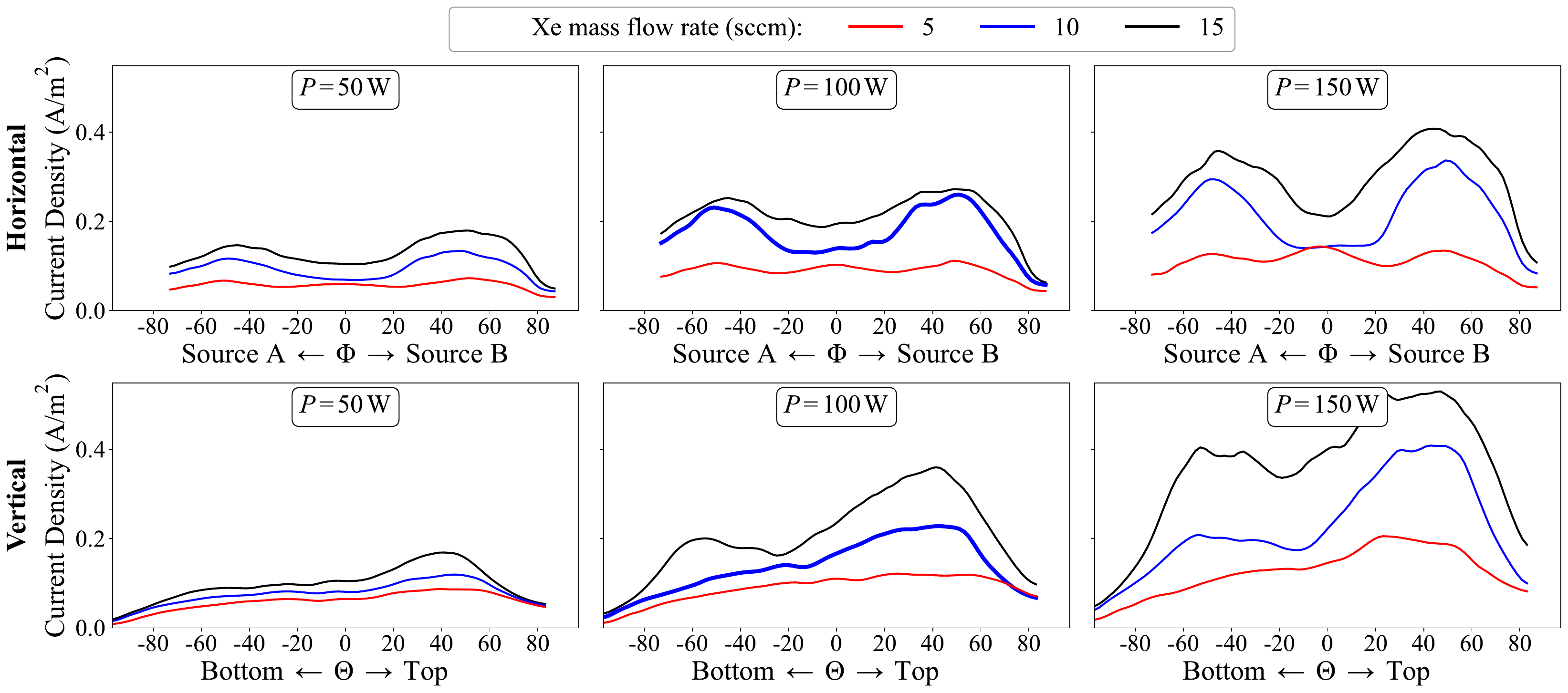}
    \caption{Ion current density angular distribution from FC, in the horizontal and vertical plane. The reference case is highlighted by a thicker line. Each curve is the average from two measurements; the mean relative error is 2.3\%.}
    \label{fig:FC all}
\end{figure}

\begin{figure}
    \centering
    \includegraphics[width=\linewidth]{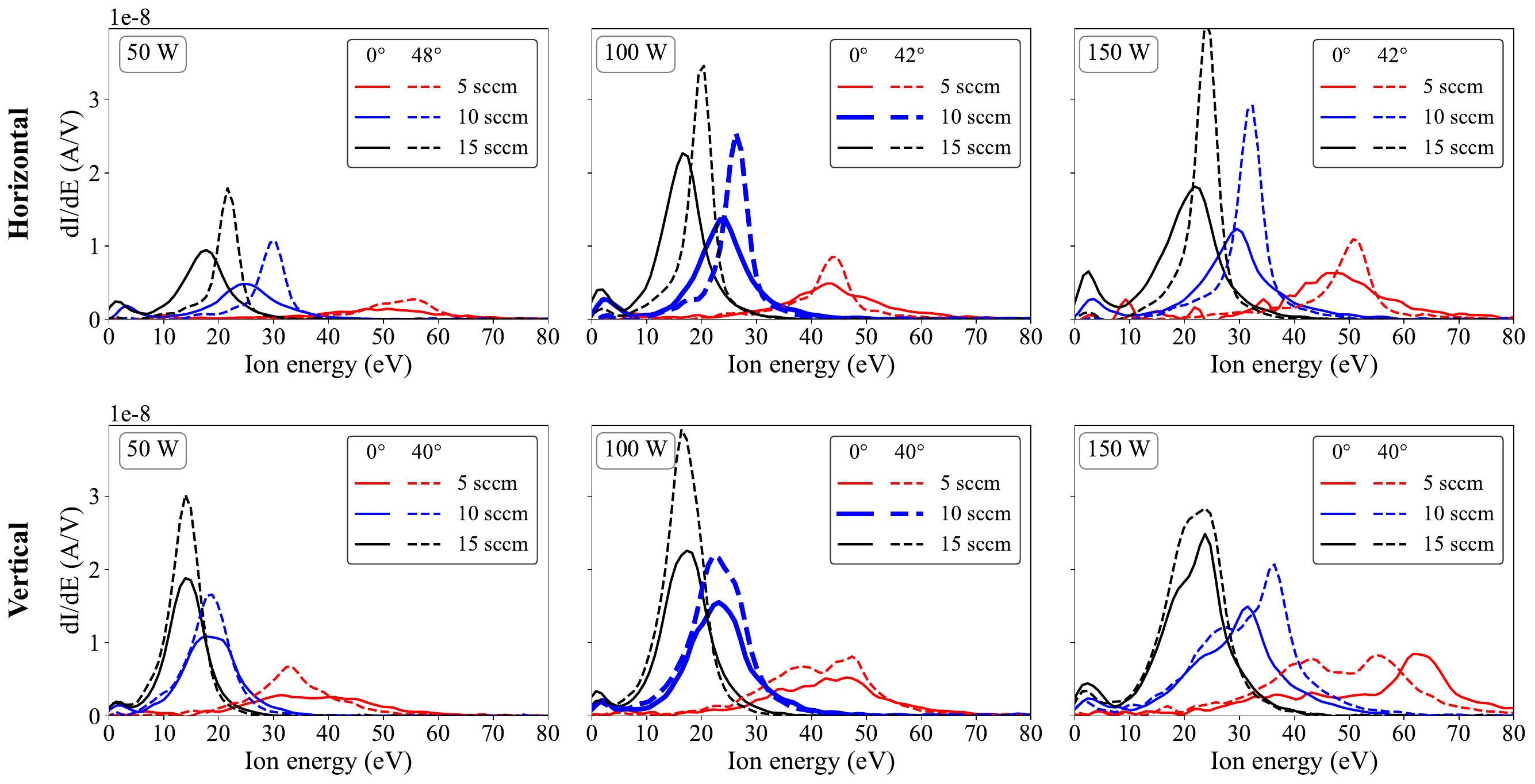}
    \caption{Ion energy distribution functions (IEDFs) from the RPA in the horizontal and vertical plane. IEDFs are acquired along the centerline (0 deg) and along the angle of maximum current density obtained by FC measurements (see figure ~\ref{fig:FC all}). The reference case is highlighted by a thicker line.}
    \label{fig:RPA all}
\end{figure}

The plasma density map also attests to the plasma expansion that takes place downstream along the $y$ axis, but there is also visible lateral expansion from each of the sources. Slow ions generated in the core arch tube, and guided by the electric field, could be responsible for these structures. Indeed, the ion current density profiles shown in figure \ref{fig:FC all} are coherent with this interpretation.

The measurements in the vertical plane
(right side of figure \ref{fig:LP_reference_combined})
depict a plasma expansion that is deflected asymmetrically sideways toward $z>0$. The intersection of the magnetic arch tube and the core tube with this plane are represented in the panels as red and yellow lines, respectively.
This asymmetry is not fully evidenced by the 
plasma potential map, which shows a relatively uniform plasma potential in the $z$ direction. Mild values of $\phi$ are found in the interior of the magnetic arch region, corresponding with those observed in the horizontal plane, with a repeatability error of approximately 20 \%.
We note that the setup had to be reassembled and repositioned after the horizontal measurements to enable the vertical ones as a possible cause for the loss of repeatability. 
Outside of the cross section of the magnetic arch, the potential fall is slightly larger toward $z>0$ than toward $z<0$:

On the other hand, the electron temperature distribution displays a clear asymmetry. High $T_e$ is observed not only within the core tube, as observed in the horizontal scans, but also extending vertically beyond its nominal boundary on one side of the plume. 
Likewise, the maximum of the density map is located at or near the $z=0$ axis.
However, it is also asymmetric and protrudes sideways in the same direction as the electron temperature, with higher densities located on the top side ($z>0$). 
All this makes it manifest that the plasma expansion in a closed magnetic arch is not fully symmetric about its fundamental plane, with electrons being able to drift transversely, and consequently, guiding the ion plume in one preferential direction.
We return to the discussion of this asymmetry and its likely causes further down \ref{sec:asymmetry}. 

The influence of the xenon mass flow rate $\dot m$ on the plasma parameters is summarized in Table~\ref{tab:plasma_summary}, where we report the value of $\phi$, $T_e$ and $n$ at three spatial points, labeled in the first plot of figure \ref{fig:LP_reference_combined}.
Increasing the mass flow rate leads to a decrease in electrostatic potential and electron temperature, consistent with the general behavior observed in ECR sources, where $
\phi$ and $T_e$ typically scale dominantly with $P/\dot m$ \cite{inchingoloPlumeCharacterizationWaveguide2023,ECRperformanceSheppard}. At 5~sccm, peak potential and electron temperatures values reach close to $40$ V and 21~eV respectively, near the exit of the sources, while at 15~sccm they decrease to approximately $14$ V and 5~eV. 
In all cases, $T_e$ remains largest within the core tube (exemplified by point $B$). 
In contrast, plasma density increases with $\dot m$, reflecting the increased availability of neutral particles for ionization, and also consistent with expected ECR source behavior in this power range. 
The maximum density shifts from being in the magnetic arch region (point $B$) to further upstream (i.e. toward point $A$) as the mass flow rate is increased.

\begin{table}[h]
\centering
\begin{tabular}{l|ccc|ccc|ccc}
\toprule
& \multicolumn{9}{c}{Mass flow rate (sccm)} \\
\cmidrule(lr){2-10}

& \multicolumn{3}{|c|}{5} 
& \multicolumn{3}{|c|}{10} 
& \multicolumn{3}{|c|}{15} \\
\cmidrule(lr){2-4} \cmidrule(lr){5-7} \cmidrule(lr){8-10}
& A & B & C & A & B & C & A & B & C \\
\midrule
$\phi$ (V) 
& 18.0 & 22.0 & 11.0 
& 7.0 & 9.0 & 7.0 
& 5.5 & 7.5 & 4.5 \\

$T_e$ (eV) 
& 10.80 & 19.04 & 5.37 
&4.52 & 10.40 & 4.28 
& 3.43 & 3.76 & 2.76 \\

$n$ ($10^{15} \mathrm{m^{-3}}$)  
& $8.93$ & $12.4$ & $2.18$
& $16.9$ & $29.5$ & $4.25$
& $23.5$ & $22.1$ & $4.77$ \\
\bottomrule
\end{tabular}
\caption{Summary of plasma parameters at $P=100$W for different xenon mass flow rates. The values are reported for each interest point A,B,C located respectively on the centerline at 50, 80, 200 mm downstream of the cluster (refer to figure \ref{fig:LP_reference_combined}).}
\label{tab:plasma_summary}
\end{table}

Faraday cup measurements provide the angular distribution of ion current density $j_i$ in the plume of the ECRT cluster; results are shown in figure ~\ref{fig:FC all} for  mass flow rate $\dot m =5,10,15$ sccm and input power $P=50,100,150$ W. Ions, being essentially unmagnetized, respond mainly to the electrostatic field, which itself results from the interplay between ions and the magnetized electrons.
Measurements in the horizontal plane reveal a structured plume composed of two dominant ion current density peaks located at approximately $\Phi \approx \pm 45$ deg, which are attributed to the ion beam emitted by each plasma source. Indeed, at the distance of measurement ($28$ cm), ion current density is significantly lower in the central region between the sources, resulting in a hollow plume topology. While hollow plumes have been reported for single-source waveguide ECR thrusters and other electrodeless plasma thrusters \cite{takao2013ecrt, charoy2015ecr, inchingolo2024thrust}, the present results suggest that this structure primarily arises from the dual-source geometry of the cluster rather than from the intrinsic divergence of each individual source. 
Increasing power or mass flow rate increases the overall value of $j_i$ and enhances the doubly-peaked structure.
Increasing microwave power leads to a nearly monotonic increase in ion current density;
However, the overall rise of $j_i$ between 5 and 10~sccm is significantly larger than between 10 and 15~sccm, suggesting a reduction of efficiency at higher neutral densities.
A small but systematic difference in value between the two peaks is noted, with the side associated with Source~B exhibiting a slightly higher current. This difference becomes more significant at higher mass flow rates. We attribute this to the small asymmetries between the sources (in magnetic field strength as discussed previously, a possible imbalance in mass flow rate between the sources and/or to tolerances in the manufacturing and assembling of the sources. 
At the lowest xenon mass flow rate setting of 5 sccm, a third central peak can be observed.
Drawing analogy from other plasma plumes, it is expected that the multi-peaked current profiles smooth out into a bell-shaped profile further downstream \cite{meri15collisionless}.

Measurements performed in the vertical plane reveal once again the non-symmetric plume expansion with respect to the horizontal plane. A hollow profile similar to the one observed in the horizontal measurements can be seen, although less pronounced. The ion current density is systematically higher towards the top side of the cluster ($z>0$), indicating a net lateral deflection of the ion exhaust, consistent with the lower $\phi$ and $n$ values on this side of the vertical plane seen in figure \ref{fig:LP_reference_combined}. Away from this peak of $j_i$, which occurs at $\Theta\simeq 40$ deg, the ion current decays monotonically, except at high powers and mass flow rates, for which a  secondary peak forms in this plane at $
\Theta \simeq (-60, -40)$ deg.

The RPA was used to compute the ion energy distribution function (IEDF) at different locations in the plume for the same power and mass flow rate settings (figure ~\ref{fig:RPA all}). Measurements were performed along the centerline and at the angles of maximum ion current density identified in the Faraday cup data(figure ~\ref{fig:FC all}), in both the horizontal and vertical planes.

For the case $P=100$ W, $\dot{m}=10$ sccm , the IEDF exhibits a clear peak at approximately 23~eV at 0 deg and 26~eV off-axis. This range of most probable ion energy remains nearly constant across measurement angle in both planes, indicating that ions experience a relatively similar acceleration potential regardless of emission direction. This is consistent with ion acceleration dominated by plasma expansion in the ambipolar electric field \cite{ahedo2011magnetic}, and is in good agreement with Langmuir probe measurements showing plasma potentials on the order of 22~V near the source region (figure  \ref{fig:LP_reference_combined}). This agreement is also true for 5 and 15 sccm with the most probable ion energy being always only a few eVs higher than the plasma potential at the source exhaust. The fact that an important portion of the ions exceed the maximum potential measured by LP, indicates that $\phi$ is higher within the sources and that a major amount of ions are produced inside the chambers. 

The IEDF shape varies spatially. Along the centerline of the magnetic arch, the distributions are broader than at the angles of maximum ion current density, where they are sharper. This suggests a wider range of ion origins (and  therefore acceleration histories) within the central arch region.

The operating parameters influence the IEDF in a manner consistent with that observed with other diagnostics. Decreasing the mass flow rate leads to significant increase in most probable ion energies and lower peak intensities, from $\simeq 17$ eV at $15$ sccm to $\simeq 45$ eV at $5$ sccm, reflecting reduced ion flux but enhanced acceleration. 
The effect of microwave power on ion energy is comparatively weaker, but important too. Increasing power leads to higher IEDF peak intensities, consistent with Faraday cup measurements, while the most probable ion energy exhibits a moderate increase from $\simeq25$ eV at $50$ W to $\simeq30$ eV at $150$ W in the horizontal plane.
This indicates that $\dot{m}$ is the primary control parameter of ion energy.

Results in the vertical plane show that ions have roughly the same distribution at both 0 deg and the angle of the current peak, hinting at a similar acceleration history for ions at both angles. 
As expected, higher intensities are measured along the direction of maximum $j_i$. 

Under low mass flow rate and high power conditions, occasional double-peaked distributions are observed (mainly in the vertical plane measurements), suggesting non-Maxwellian ion populations and possibly multiple acceleration regions within the plume.
Additionally, a low energy population of background ions can be observed in every condition, at approximately 2-3 eV. 

\subsection{Plume asymmetry and electron drifts}\label{sec:asymmetry}

\begin{figure}
    \centering
    \includegraphics[width=.75\linewidth]{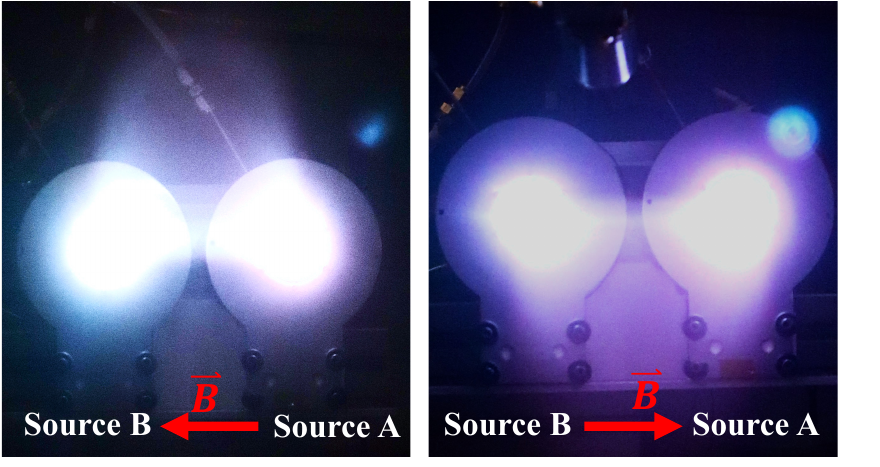}
    \caption{Photographs of the plasma plume from the front ($P=100$W, $\dot{m}=10$ sccm). Nominal magnetic field orientation (left) and reversed orientation (right). The blue circles on the top right of source B are camera lenses reflections.}
    \label{fig:asymmetry pictures}
  \end{figure}

The asymmetry in the vertical plane observed above merits its own discussion. 
Previous work with a partially-open magnetic arch did not observe any major asymmetry, although specific measurements to detect it were not carried out \cite{Boyé_Navarro-Cavallé_Merino_2025}.
In our present closed arch experiment, a pronounced and persistent asymmetry of the exhaust plume is visibly observed  across all investigated operating conditions, demonstrating that the plasma expansion is not symmetric about the horizontal axis.
A photograph showing this deflection is shown in figure \ref{fig:asymmetry pictures}(left).

Figure \ref{fig:Bfield} shows that a small difference in magnetic field $\bm B$ exists between the two sources. It is also likely that small differences in the power and mass flow rate reaching each source exist.
While these differences are likely responsible for the unequal $j_i$ peaks observed in the horizontal plane in figure \ref{fig:FC all},
it is not reasonable to attribute the lateral deflection in the vertical plane to them.
Instead, the curved field lines in the magnetic arch and the direction of the $\bm B$ vector are suspected to partake in this sideways displacement of the plume.
Since only the electrons are magnetized and thus responsive to $\bm B$, we direct our attention to the magnetic field geometry and direction and their effects on the electron dynamics.



\begin{figure}
    \centering
    \includegraphics[width=\linewidth]{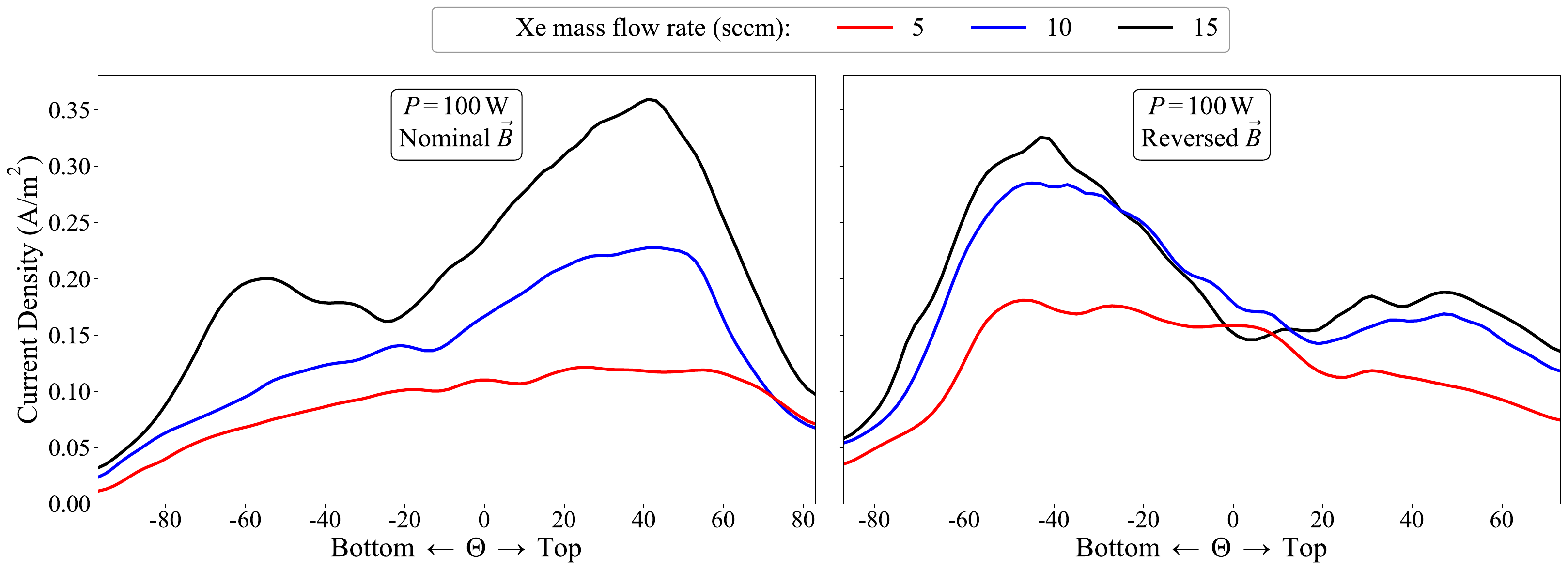}
    \caption{Ion current density distribution in the vertical plane at $P=100$W, with the nominal magnetic field (left) and reversed polarity (right).}
    \label{fig:FC asymmetry}
  \end{figure}

The magnetic origin of the asymmetry is demonstrated by reversing the polarity of the magnetic arch. This is accomplished by reversing the polarity of the permanent magnets and coils accordingly. The permanent magnets were dismounted and reversed, while the coils wiring was simply inverted.  
As illustrated in figure \ref{fig:asymmetry pictures}(right), reversing $\bm B$ also changes the direction of deflection of 
the plume from top to bottom. 
This change is also observed in the additional  Faraday cup measurements of figure \ref{fig:FC asymmetry}.  

The behavior with the asymmetry with field reversal hints at electron drifts as a possible explanation for the lateral deflection, as the degree of ion magnetization is small already at the source (measured by the dimensionless ion gyrofrequency at the source exit, $\hat\Omega_{i0}=eB_0R/\sqrt{T_{e0} m_i}<1$, where the characteristic length $R$ is taken to be the radius of the sources), so ion magnetic forces are known to play a small role in the expansion.
This is best understood by analyzing the gyrocenter drifts of an individual electron located somewhere on the centerline of the arch, around points A or B in figure \ref{fig:LP_reference_combined}.
To first order in the electron Larmor radius, these
include the $\Vec{E}\times \Vec{B}$ drift, the $\nabla \Vec{B}$ drift, and  the curvature drift:
\begin{align}
\Vec{v}_{E} &= \frac{\bm{E}\times\bm{B}}{B^2};
&
\bm{v}_{g} &= \frac{m v_\perp^2}{2 e B^2} \, \bm{B}\times\nabla \ln B;
&
\bm{v}_{c} &= \frac{m v_\parallel^2}{e B^2} \, \bm{B}\times\bm{\kappa},
\end{align}
where $\bm{\kappa} = (\bm{B}/B\cdot\nabla)\bm{B}/B$ is the curvature vector of the magnetic field. Assuming isotropic thermal electrons with 
$v_{\parallel e} = v_{\perp e} = \sqrt{T_e/m_e} \simeq  10^6$ m/s (corresponding to $T_e\simeq 6$ eV as a representative value in the magnetic arch), an axial electric field around $E_y\simeq -100$ V/m, a curvature radius and a gradient length of $1/\kappa \simeq 50$ mm, and a magnetic field $B\simeq 100$ G,
we can estimate $v_E\simeq v_g\simeq v_c\simeq10^4$ m/s and all of them pointing in the $+z$ direction for the nominal configuration ($\bm B$ pointing from source A to source B). 
Computing the total drift velocities along $z$ at the points of interest A and B (see table \ref{tab:plasma_summary} and figure \ref{fig:LP_reference_combined}a), yields 5.2 km/s and 22.5 km/s respectively.
Moreover, while $v_E$ may switch directions after the core arch tube, where $E_y>0$, the other two always point in the $+z$ direction.
While this is a small drift velocity for thermal electrons (their tangent $v_z/v_\parallel$ gives about 0.28 deg for point A and 0.72 deg for point B), trapped electrons with comparatively large residence time can easily be deflected along the $z$ by this effect.
More importantly, the direction of the electron drifts switches to the $-z$ direction when the magnetic field is reversed. As the electrons drive the ions through the ambipolar field, the electron drifts provide a consistent explanation for the asymmetric ion current profiles observations.

Incidentally, and while outside of the scope of this work, 
thrust vectoring in the magnetic arch configuration merits a brief mention. 
Thrust vector control in the $x$ axis can be pursued by applying differential power / mass flow rate / field strength to each of the two plasma sources. 
In addition, the existence of the lateral deflection described in this section, adds the possibility of controlling it with the amplitude and direction of the applied field, open the intriguing possibility of exploiting this mechanism for thrust vectoring in the $z$ axis.


\section{Conclusions}\label{sec:conclu}

This work provides a detailed experimental characterization of the plume of a clustered electron cyclotron resonance thruster operating in a closed magnetic arch configuration. Using complementary diagnostics, the plume structure was fully resolved in two orthogonal planes (horizontal and vertical), providing, to the authors’ knowledge, the first comprehensive two-plane, two-dimensional mapping of such a configuration.

The results reveal a structured and magnetically controlled plume, in which ion acceleration is governed by ambipolar expansion, while electron confinement within the magnetic arch plays a central role in ion production and shaping the electrostatic field. 
A magnetic tube of high electron temperature is present in the core region of the magnetic arch, which likely drives the external ionization of the propellant in our setup, which features two sources with low propellant utilization.
Plasma density, ion current, and ion energy measurements 
indicate that the expansion in the horizontal plane is generally two-peaked and has a relatively ample half-angle.
A pronounced and reproducible plume asymmetry is observed across all diagnostics, wherein the plasma jet is deflected away from the central horizontal plane in the direction dictated by the electron gyrocenter drifts. Its magnetic origin is clearly established, and drift-driven electron transport within the arch has been identified as the likely  mechanism.

These findings highlight the strong coupling between magnetic topology and plume properties, and open new perspectives for active control of plasma exhaust in electrodeless thrusters. Future work should investigate the influence of magnetic arch topology, including open versus closed configurations, source tilting, and fine control of the magnetic field using additional coils, in order to better understand and exploit plume shaping and asymmetry control.

\paragraph{\textbf{Acknowledgments}}

This work has received funding from the European Research Council (ERC) under the European Union’s Horizon 2020 research and innovation programme (Starting Grant project ZARATHUSTRA, grant agreement No 950466). 

\bibliographystyle{unsrt}
\bibliography{bibliography,ep2}

@article{meri15collisionless,
  title = {A Collisionless Plasma Thruster Plume Expansion Model},
  author = {Merino, Mario and Cichocki, Filippo and Ahedo, Eduardo},
  year = 2015,
  month = apr,
  journal = {Plasma Sources Science and Technology},
  volume = {24},
  number = {3},
  pages = {035006},
  issn = {0963-0252, 1361-6595},
  doi = {10.1088/0963-0252/24/3/035006}
}

@article{takao2013ecrt,
  author  = {Takao, Y. and Shinohara, S.},
  title   = {Ion acceleration and thrust performance of an electrodeless plasma thruster},
  journal = {Plasma Sources Science and Technology},
  volume  = {22},
  number  = {4},
  pages   = {045006},
  year    = {2013},
  doi     = {10.1088/0963-0252/22/4/045006}
}

@article{charoy2015ecr,
  author  = {Charoy, A. and Dudeck, M.},
  title   = {Experimental study of an ECR plasma thruster},
  journal = {Acta Astronautica},
  volume  = {108},
  pages   = {1--12},
  year    = {2015},
  doi     = {10.1016/j.actaastro.2014.11.030}
}

@article{ahedo2011magnetic,
  author  = {Ahedo, E. and Merino, M.},
  title   = {Magnetic nozzle effects on plasma acceleration},
  journal = {Physics of Plasmas},
  volume  = {18},
  number  = {5},
  pages   = {053504},
  year    = {2011},
  doi     = {10.1063/1.3587084}
}

@inproceedings{boye:hal-04686118,
  TITLE = {{Determining ion velocity in the interconnected plume of a cluster of two ECRTs using 2D LIF}},
  AUTHOR = {Boy{\'e}, C{\'e}lian and Navarro-Cavall{\'e}, Jaume and Merino, Mario and Lecervoisier, Alexis and Mazouffre, St{\'e}phane},
  URL = {https://hal.science/hal-04686118},
  BOOKTITLE = {{Electric Rocket Propulsion Society - IEPC 2024 proceedings}},
  ADDRESS = {Toulouse, France},
  SERIES = {Electric Rocket Propulsion Society - IEPC 2024 proceedings},
  PAGES = {IEPC 2024-717},
  YEAR = {2024},
  MONTH = Jun,
  HAL_ID = {hal-04686118},
  HAL_VERSION = {v1}
}

@article{Boyé_Navarro-Cavallé_Merino_2025, title={Ion current and energy in the Magnetic Arch of a cluster of two ECR plasma sources}, volume={4}, DOI={10.1007/s44205-025-00100-w}, number={1}, journal={Journal of Electric Propulsion}, author={Boyé, C. and Navarro-Cavallé, J. and Merino, M.}, year={2025}, month={Feb}}

@article{Guaita_2025,
doi = {10.1088/1361-6595/adab8e},
url = {https://dx.doi.org/10.1088/1361-6595/adab8e},
year = {2025},
month = {jan},
publisher = {IOP Publishing},
volume = {34},
number = {1},
pages = {015007},
author = {Guaita, M and Ahedo, E and Merino, M},
title = {PIC/Fluid simulations of the plasma expansion in a planar magnetic arch},
journal = {Plasma Sources Science and Technology}
}

@article{Lobbia_Beal_2017, title={Recommended practice for use of Langmuir probes in electric propulsion testing}, volume={33}, DOI={10.2514/1.b35531}, number={3}, journal={Journal of Propulsion and Power}, author={Lobbia, Robert B. and Beal, Brian E.}, year={2017}, month={May}, pages={566–581}}

@article{inchingolo2024thrust,
  title={Thrust measurements of a waveguide electron cyclotron resonance thruster},
  author={Inchingolo, MR and Merino, Mario and Wijnen, Mick and Navarro-Cavall{\'e}, Jaume},
  journal={Journal of Applied Physics},
  volume={135},
  number={9},
  year={2024},
  publisher={AIP Publishing}
}

@article{cannat2015optimization,
  title={Optimization of a coaxial electron cyclotron resonance plasma thruster with an analytical model},
  author={Cannat, F and Lafleur, T and Jarrige, J and Chabert, P and Elias, P-Q and Packan, D},
  journal={Physics of Plasmas},
  volume={22},
  number={5},
  year={2015},
  publisher={AIP Publishing}
}

@article{inchingoloPlumeCharacterizationWaveguide2023,
  title = {Plume Characterization of a Waveguide {{ECR}} Thruster},
  author = {Inchingolo, M. R. and Merino, M. and {Navarro-Cavall{\'e}}, J.},
  year = 2023,
  month = mar,
  journal = {Journal of Applied Physics},
  volume = {133},
  number = {11},
  publisher = {AIP Publishing},
  issn = {0021-8979},
  doi = {10.1063/5.0138780},
  urldate = {2026-03-11},
  langid = {english}
}

@article{merinoPlasmaAccelerationMagnetic2023,
  title = {Plasma Acceleration in a Magnetic Arch},
  author = {Merino, Mario and {Garc{\'i}a-Lahuerta}, Diego and Ahedo, Eduardo},
  year = 2023,
  month = jun,
  journal = {Plasma Sources Science and Technology},
  volume = {32},
  number = {6},
  pages = {065005},
  issn = {0963-0252, 1361-6595},
  doi = {10.1088/1361-6595/acd476},
  urldate = {2025-06-19},
  langid = {english}
}

@incollection{vereenCharacterizationClusterHigh2017,
  title = {Characterization of a {{Cluster}} of {{High Power Helicon Thrusters}}},
  booktitle = {53rd {{AIAA}}/{{SAE}}/{{ASEE Joint Propulsion Conference}}},
  author = {Vereen, Keon and Kimber, Angela and Correy, John and Olson, David and Martin, Hans and Winglee, Robert},
  year = 2017,
  month = jul,
  series = {{{AIAA Propulsion}} and {{Energy Forum}}},
  publisher = {{American Institute of Aeronautics and Astronautics}},
  doi = {10.2514/6.2017-4628},
  urldate = {2026-03-11}
}

@article{difedeMagneticNozzlePerformance2023,
  title = {Magnetic Nozzle Performance in a Cluster of Helicon Plasma Thrusters},
  author = {Di Fede, Simone and Manente, Marco and Jo{\~a}o Comunian, Paolo and Magarotto, Mirko},
  year = 2023,
  month = jun,
  journal = {Plasma Sources Science and Technology},
  volume = {32},
  number = {6},
  pages = {065013},
  publisher = {IOP Publishing},
  issn = {0963-0252},
  doi = {10.1088/1361-6595/acdaf2},
  urldate = {2026-03-11},
  langid = {english}
}

@article{littleIonizationCurrentSheet2022,
  title = {Ionization and Current Sheet Formation in Inductive Pulsed Plasma Thrusters},
  author = {Little, Justin M. and McCulloh, Gordon I. and Marsh, Cameron},
  year = 2022,
  month = sep,
  journal = {Journal of Applied Physics},
  volume = {132},
  number = {9},
  pages = {093301},
  issn = {0021-8979},
  doi = {10.1063/5.0102077},
  urldate = {2026-03-11}
}

@article{navarro-cavalleExperimentalCharacterization12018,
  title = {Experimental Characterization of a 1~{{kW Helicon Plasma Thruster}}},
  author = {{Navarro-Cavall{\'e}}, J. and Wijnen, M. and Fajardo, P. and Ahedo, E.},
  year = 2018,
  month = mar,
  journal = {Vacuum},
  volume = {149},
  pages = {69--73},
  issn = {0042-207X},
  doi = {10.1016/j.vacuum.2017.11.036},
  urldate = {2026-03-11}
}

@article{peterschmittImpactMicrowaveCoupling2021,
  title = {Impact of the {{Microwave Coupling Structure}} on an {{Electron-Cyclotron Resonance Thruster}}},
  author = {Peterschmitt, Simon and Packan, Denis},
  year = 2021,
  month = nov,
  journal = {Journal of Propulsion and Power},
  volume = {37},
  number = {6},
  pages = {806--815},
  publisher = {{American Institute of Aeronautics and Astronautics}},
  issn = {0748-4658},
  doi = {10.2514/1.B38156},
  urldate = {2026-03-16}
}

@article{ECRperformanceSheppard,
author = {Sheppard, Anna J. and Little, Justin M.},
title = {Performance Analysis of an Electron Cyclotron Resonance Thruster with Various Propellants},
journal = {Journal of Propulsion and Power},
volume = {38},
number = {6},
pages = {998-1008},
year = {2022},
doi = {10.2514/1.B38698},

URL = { 
    
        https://doi.org/10.2514/1.B38698
},
eprint = { 
    
        https://doi.org/10.2514/1.B38698
}

}

@article{ahedo2010,
  title = {Two-Dimensional Supersonic Plasma Acceleration in a Magnetic Nozzle},
  author = {Ahedo, E. and Merino, M.},
  year = 2010,
  month = jul,
  journal = {Physics of Plasmas},
  volume = {17},
  number = {7},
  eprint = {https://pubs.aip.org/aip/pop/article-pdf/doi/10.1063/1.3442736/15646200/073501\_1\_online.pdf},
  pages = {073501},
  issn = {1070-664X},
  doi = {10.1063/1.3442736}
}

@article{TakahashiMN,
  title = {Electron Diamagnetic Effect on Axial Force in an Expanding Plasma: Experiments and Theory},
  author = {Takahashi, Kazunori and Lafleur, Trevor and Charles, Christine and Alexander, Peter and Boswell, Rod W.},
  journal = {Phys. Rev. Lett.},
  volume = {107},
  issue = {23},
  pages = {235001},
  numpages = {4},
  year = {2011},
  month = {Nov},
  publisher = {American Physical Society},
  doi = {10.1103/PhysRevLett.107.235001},
  url = {https://link.aps.org/doi/10.1103/PhysRevLett.107.235001}
}

@article{TakahashiHPT,
author = {Takahashi, Kazunori},
year = {2019},
month = {12},
pages = {},
title = {Helicon-type radiofrequency plasma thrusters and magnetic plasma nozzles},
volume = {3},
journal = {Reviews of Modern Plasma Physics},
doi = {10.1007/s41614-019-0024-2}
}

@article{BoswellHPT2,
title = "Laboratory evidence of supersonic ion beam generated by a current-free {"}helicon{"} double-layer",
author = "C. Charles and Boswell, \{R. W.\}",
year = "2004",
month = apr,
doi = "10.1063/1.1652058",
language = "English",
volume = "11",
pages = "1706--1714",
journal = "Physics of Plasmas",
issn = "1070-664X",
publisher = "American Institute of Physics",
number = "4",
}

@article{MillerECRT,
author = {MILLER, DAVID B. and GIBBON, EDWARD F.},
title = {Experiments with an electron cyclotron resonance plasma accelerator},
journal = {AIAA Journal},
volume = {2},
number = {1},
pages = {35-41},
year = {1964},
doi = {10.2514/3.2210},

URL = { 
    
        https://doi.org/10.2514/3.2210},
eprint = { 
    
        https://doi.org/10.2514/3.2210
}
}

@phdthesis{Sercel_Culick_1993, place={Pasadena, Calif}, title={An experimental and Theoretical Study of the ECR Plasma Engine}, journal={An experimental and theoretical study of the ECR plasma engine}, school={California Institute of Technology}, author={Sercel, Joel Christopher and Culick, F.}, year={1993}}

@article{jime23,
    title = {Analysis of a cusped helicon plasma thruster discharge},
    volume = {32},
    issn = {0963-0252, 1361-6595},
    url = {https://iopscience.iop.org/article/10.1088/1361-6595/ad01da},
    doi = {10.1088/1361-6595/ad01da},
    language = {en},
    number = {10},
    urldate = {2024-12-19},
    journal = {Plasma Sources Science and Technology},
    author = {Jiménez, Pedro and Zhou, Jiewei and Navarro-Cavallé, Jaume and Fajardo, Pablo and Merino, Mario and Ahedo, Eduardo},
    month = oct,
    year = {2023},
    note = {3 citations (Semantic Scholar/DOI) [2024-12-21]},
    pages = {105013},
}

@STRING{de  = {Diego Escobar}}

@STRING{mr  = {Mercedes Ruiz}}

@STRING{and = { and }}

@STRING{psst = {Plasma Sources Science and Technology}}

@STRING{iepc = {International Electric Propulsion Conference}}

@STRING{AIAA = {American Institute of Aeronautics and Astronautics}}

@STRING{iop = {IOP Publishing}}

@misc{EP2,
  title = {Equipo de Propulsion Espacial y Plasmas},
  howpublished = {\url{http://ep2.uc3m.es}}
}

@article{boye25a,
    title = {Ion current and energy in the magnetic arch of a cluster of two {ECR} plasma sources},
    volume = {4},
    issn = {2731-4596}, 
    doi = {10.1007/s44205-025-00100-w},   
    number = {1},
    journal = {Journal of Electric Propulsion},
    author = {C\'elian Boy\'e and Jaume Navarro-Cavall\'e and Mario Merino},
    month = feb,
    year = {2025}, 
    pages = {10},
}

@article{perr24b,
  title = {Far-{F}ield {P}lume {C}haracterization of a {L}ow-{P}ower {C}ylindrical {{Hall}} {T}hruster},
  author = {Perrotin, Tatiana and Vinci, Alfio E. and Mazouffre, St{\'e}phane and Fajardo, Pablo and Ahedo, Eduardo and {Navarro-Cavall{\'e}}, Jaume},
  year = {2024},
  month = jul,
  journal = {Journal of Applied Physics},
  volume = {136},
  number = {4},
  pages = {043304},
  doi = {10.1063/5.0207003},
  publisher={AIP Publishing LLC}
}

@article{inch24a,
    author = {Inchingolo, M. R. and M. Merino and M. Wijnen and J. Navarro-Cavall\'e},
    title = {{Thrust Measurements of a Waveguide Electron Cyclotron Resonance Thruster}},
    journal = {Journal of Applied Physics},
    volume = {135},
    number = {9},
    year={2024},
    month = {3},
    url = {https://doi.org/10.1063/5.0186778},
    doi = {10.1063/5.0186778},
    publisher={AIP Publishing LLC}
}

@article{svil23,
author={\'A. S\'anchez-Villar and F. Boni and V. D\'esangles and J. Jarrige and D. Packan and E. Ahedo and M. Merino},
title={Comparison of a hybrid model and experimental measurements for a dielectric-coated coaxial {ECR} thruster},
journal = psst,
doi = {10.1088/1361-6595/acb00c},
year = {2023},
publisher = {IOP Publishing},
volume = {32},
number = {1},
pages = {014002},
}

@article{svil21a,
  doi = {10.1088/1361-6595/abde20},
  author = {A. S\'anchez-Villar and J. Zhou and M. Merino and E. Ahedo},
  title = {Coupled plasma transport and electromagnetic wave simulation of an {ECR} thruster},
  journal = psst,
  volume = {30},
  number = {4},
  pages = {045005},
  ISSN = {1089-7674},
  year = 2021,
  publisher = {{IOP} Publishing}
}

@inproceedings{inch23b,
  title={Faraday Cup Design for Electrodeless Plasma Thrusters},
  author={M. Inchingolo and J. Navarro-Cavall\'e},
  booktitle={10th EUCASS Conference},
  year={2023},
  address = {Lausanne, Switzerland, July 9–13}
}

\end{document}